\documentclass[aps,prd,twocolumn,showpacs,superscriptaddress,groupedaddress]{revtex4}  
\usepackage{graphicx}  
\usepackage{dcolumn}   
\usepackage{bm}        
\usepackage{amssymb}   

\usepackage{subfigure}
\usepackage{epsfig}
\usepackage[tbtags]{amsmath}
\usepackage{exscale}

\newcommand{\md}{{\mathrm{d}}}

\begin{document}

\title{Zero modes and divergence of entanglement entropy}

\author{Krishnanand Mallayya} \email{krishnand@iisertvm.ac.in}
\affiliation{School of Physics, Indian Institute of Science Education
  and Research, CET Campus, Trivandrum 695016, India}

\author{Rakesh Tibrewala} \email{rtibs@iisertvm.ac.in}
\affiliation{School of Physics, Indian Institute of Science Education
  and Research, CET Campus, Trivandrum 695016, India}

\author{S. Shankaranarayanan} \email{shanki@iisertvm.ac.in}
\affiliation{School of Physics, Indian Institute of Science Education
  and Research, CET Campus, Trivandrum 695016, India}

\author{T. Padmanabhan}
\email{paddy@iucaa.ernet.in}
\affiliation{IUCAA, Pune University Campus, Ganeshkhind, Pune 411007, INDIA}

\begin{abstract}
We investigate the cause of the divergence of the entanglement entropy 
for the free scalar fields in $(1+1)$ and $(D + 1)$ dimensional
space-times. In a canonically equivalent set of variables, we show explicitly that
the divergence in the entanglement entropy of the continuum field in $(1 + 1)-$ dimensions is 
due to the accumulation of large number of near-zero frequency modes as 
opposed to the commonly held view of divergence having UV origin. The feature 
revealing the divergence in 
zero modes is related to the observation that the entropy is invariant 
under a hidden scaling transformation even when the Hamiltonian is not. 
We discuss the role of dispersion relations and the dimensionality of 
the space-time on the behavior of entanglement entropy. 
\end{abstract}

\maketitle
\section{Introduction}

Entanglement entropy, a popular measure to quantify quantum entanglement, has
become a subject of intensive theoretical investigation, especially
for systems with many degrees of freedom \cite{CardyQFT, CardyCFT,
  WilczekCFT} and is being used to characterize properties of a wide
spectrum of systems like quantum information processing \cite{13, 3},
quantum phase transition \cite{sondhi,
  sachdev} and entropy of black holes \cite{Bombelli, srednicki,
  SolodukhinReview, ShankiReview}.

However, the entanglement entropy of free quantized fields is found to
be divergent \cite{Bombelli, WilczekGeometricEntropy} and some form of
regularization has to be used in order to extract useful information
from it (also see \cite{SoloNestUVmodified, PaddyZeroPointArea,
  SoloNestShortDistReg} in this context). Furthermore, in general, it
is difficult to get an analytic handle on entanglement entropy except
for a few special cases like $(1+1)$ dimensional CFTs
\cite{CardyCFT} \footnote{It is important to note that the analytical 
expression for entanglement entropy uses the Replica trick 
\cite{WilczekCFT}}. 

In this work,  with the aim to gain a better
understanding of the divergence of entanglement entropy as well as to
have better analytic control, we consider in detail the entanglement entropy of
free scalar field regularized on a spatial lattice in $(1+1)$-dimensional space-time. 
With the insights  gained in $(1 + 1)-$ dimensions, we extend the results to higher 
dimensions.   

Specifically, we obtain analytical expression for the entanglement entropy 
by tracing over a single oscillator of the lattice regularized scalar field. 
The analytical expression provides two interesting features which, to our knowledge, 
have not been noted in the literature: 
(i) entanglement entropy is invariant under a scaling transformation even when 
the Hamiltonian is not, and 
(ii) the divergence in entanglement entropy in $(1 +1)-$ dimensions in the continuum limit is 
due to the presence of a large number of near zero modes (and 
is not of UV origin as commonly believed). In the case of higher 
dimensions, accumulation of zero modes occur, however,  the entropy remains finite (non-divergent). 

The work is organized as follows. In the next section we start with
the discretized version of free, massive scalar field in $(1+1)$
dimensions and define the covariance matrix for the corresponding
Hamiltonian and show its relation to the entanglement entropy. In
section (3), after performing a canonical transformation on the phase
space of the scalar field theory, we show that the divergence of
entanglement entropy in the continuum is due to the presence of near 
zero frequency modes. It is further shown that the entropy is divergent
even when a single oscillator is traced over. Section (4) is devoted to
regulating this divergence using a suitable infra-red cut-off. 
In section (5) we consider the effect of spatial dimensions on entanglement 
entropy of free fields.  We conclude in section (6). We set $c=\hbar=1$.

\section{Entanglement entropy and covariance matrix in $(1 + 1)-$ dimensions}

The system of interest is the $(1+1)$ dimensional massive, free scalar
field theory described by the Lagrangian
\begin{equation} \label{scalar field lagrangian}
L^{\rm (1D)} = \frac{1}{2}\int \md
x\left(\dot{\phi}^{2}-(\nabla\phi)^{2}-m_{f}^{2}\phi^{2}\right),
\end{equation}
where $m_{f}$ is the mass of the scalar field. The corresponding
Hamiltonian is
\begin{equation} \label{scalar field hamiltonian in continuum}
H^{\rm (1D)}=\frac{1}{2}\int \md
x\left(\pi^{2}+(\nabla\phi)^{2}+m_{f}^{2}\phi^{2}\right).
\end{equation}
As mentioned earlier, the ground state entanglement entropy for such a
system is divergent. In order to gain better understanding of the divergence, 
we place the system on a spatial lattice with lattice spacing $a$. Using the
notation $\phi_{n}=\phi(x_{n})$ $(n\in \mathbb{I})$, where $x_{n}=an$
denotes the position of the lattice points, the discretized Hamiltonian is
\begin{equation} \label{discretized hamiltonian for scalar field}
H^{\rm (1D)} = \frac{1}{2}\sum_{n=0}^{N-1}\left(\frac{\pi_{n}^{2}}{a}
+\frac{1}{a}(\phi_{n+1}-\phi_{n})^{2}+am_{f}^{2}\phi_{n}^{2}\right),
\end{equation}
where $\pi_{n}=a\dot{\phi}_{n}$ is the canonically conjugate momentum 
of $\phi_{n}$. It is important to note that we have an
infinite lattice in mind so that $N\rightarrow\infty$; the continuum
limit corresponds to $a\rightarrow0$. In the following we assume
   periodic boundary conditions $\phi_{0}=\phi_{N}$ (the final results
are, however, independent of the specific choice of boundary
conditions).

Our aim is to analytically calculate the entanglement entropy of the 
ground state wave-function of the above Hamiltonian obtained by tracing 
over first $m<N$ oscillators. This can be done by finding the ground state
wave-function and then performing the partial trace. However, for
greater generality, instead of taking this route, we calculate the
entanglement entropy by finding the covariance matrix, which for the
Hamiltonian \eqref{discretized hamiltonian for scalar field} is the
following $2N\times2N$ matrix \cite{3}
\begin{eqnarray} \label{covariance matrix}
\sigma &=& \frac{1}{2} \left[ \begin{array}{cc}
(aV)^{-1/2} & 0 \\
0 &(aV)^{1/2} \end{array} \right].
\end{eqnarray}
In the above equation $\omega_{k}$ are the normal mode frequencies
\begin{equation} \label{normal mode frequencies for discretized scalar field}
\omega_{k}^{2}=m_{f}^{2}+\frac{4}{a^{2}}\sin^{2}\left(\frac{\pi k}{N}\right), \, k=0, 1,...., N-1
\end{equation}
and $V$ is the potential matrix of the Hamiltonian \eqref{discretized
  hamiltonian for scalar field}. The matrix elements of the `position'
correlation $\Delta\phi=\frac{1}{2}(aV^{-\frac{1}{2}})$ and `momentum'
correlation $\Delta\pi =\frac{1}{2}(aV^{\frac{1}{2}})$ depend only on the
separation $(i-j)$ between the $i$th and the $j$th oscillators.

The reduced state obtained after tracing over $m (< N)$ oscillators
can be characterized from its covariance matrix $\sigma_{red}$ by
picking appropriate elements from the total matrix. 
The entanglement entropy is given by \cite{7} 
\begin{eqnarray}\label{entropy of the reduced state}
S_{m}(\rho_{red})&=&\sum_{k=1}^m
\left(\alpha_k+\frac{1}{2}\right)\log
\left(\alpha_k+\frac{1}{2}\right) \nonumber \\ &&
-\left(\alpha_k-\frac{1}{2}\right)\log \left(
\alpha_k-\frac{1}{2}\right).
\end{eqnarray}
where $\alpha_k$ are the symplectic eigen values of the reduced covariance
matrix. 

To have better analytic control in order to identify the scaling symmetry, we consider the 
simplest case of the single oscillator reduced system $(m=1)$, for which the covariance matrix is
\begin{equation} \label{single oscillator reduced state}
\sigma_{red} =\frac{1}{2N} \left[ \begin{array}{cc}
\sum_{i} \frac{1}{a\omega_{i}} & 0 \\
0 &\sum_{j} a\, \omega_{j} \end{array} \right].
\end{equation}

Here we would like to note a couple of things regarding the  
covariance matrix \eqref{single oscillator reduced state}. The $1-1$ ($2-2$)
  element of the covariance matrix, often referred to in the literature as the 
 `position' (momentum) covariance \cite{2012-Olivares-EPJ}, is 
\begin{eqnarray} 
\label{position covariance 1+1 dim}
\Delta\phi & =& \frac{1}{2N}\sum_{i=0}^{N-1}\frac{1}{\sqrt{a^{2}m_f^2+4\left(\sin^2\left(\displaystyle\frac{\pi i}{N}\right)\right)}} \\
 \label{momentum covariance 1+1 dim}
\Delta\pi &=& \frac{1}{2N}\sum_{i=0}^{N-1}\sqrt{a^{2}m_f^2+4\left(\sin^2\left(\displaystyle\frac{\pi i}{N}\right)\right)}
\end{eqnarray} 
In the continuum limit ($a\rightarrow0$), the position covariance diverges while the momentum covariance is finite (non-zero value). 
Hence, the product $\Delta\phi\Delta\pi$, which is the determinant of the covariance matrix, diverges.

This determinant
\begin{widetext}
\begin{equation} \label{determinant reduced state}
{Det(\sigma_{red})}=\frac{1}{4N^{2}}\sum_{i=0}^{N-1}\frac{1}{\sqrt{m_{f}^{2}+\frac{4}{a^{2}}\sin^{2}\left(\frac{\pi i}{N}\right)}}\sum_{j=0}^{N-1}\sqrt{m_{f}^{2}+\frac{4}{a^{2}}\sin^{2}\left(\frac{\pi j}{N}\right)}
\end{equation}
\end{widetext}
is the eigenvalue $(\alpha_1)^{2}$, from which the entropy can be calculated using \eqref{entropy of the reduced state}. 

The following points are worth noting regarding the above result: (i)
the entropy is invariant under the scaling transformations
\begin{eqnarray} \label{scaling transformation}
m_{f} &\rightarrow& \xi m_{f} \nonumber \\
a &\rightarrow& \xi^{-1}a,
\end{eqnarray}
and (ii) in the continuum limit, $a\rightarrow0$,
$\omega^{2}_{k}\rightarrow\omega^{2}(k)=m_{f}^{2}+k^{2}$ and, with the
summations going over to integrals, the entanglement entropy
$S(\rho_{red})\rightarrow \infty$ since the numerator in
\eqref{determinant reduced state} diverges for
$k\rightarrow\infty$. This is the familiar UV-divergence of the
entanglement entropy in the continuum.

In the next section, we show that the canonical transformation of the 
variables $(\phi, \pi)$ which accounts for the scaling symmetry will 
lead to a Hamiltonian that can be separated into a scale invariant
and a scale dependent part. The entanglement entropy of the resultant
Hamiltonian leads to a new way of identifying the cause of the divergence. 

\section{Zero frequency modes and divergence of entanglement entropy}

To take into account the scaling symmetry, we introduce the following
canonical rescaling of the Hamiltonian \eqref{discretized hamiltonian for
  scalar field}
\begin{equation} \label{rescaling the canonical variables}
\pi_{n}=\bar{\pi}_{n}(2+a^{2}m_{f}^{2})^{1/4}, \quad \text{and} \quad \phi_{n}=\frac{\bar{\phi}_{n}}{(2+a^{2}m_{f}^{2})^{1/4}}.
\end{equation}
In terms of these variables the Hamiltonian can be written as
\begin{equation} \label{discretized hamiltonian in terms of rescaled variables}
H^{\rm (1D)}=\frac{E_{0}}{2}\sum_{n=0}^{N-1}\left(\bar{\pi}_{n}^{2}+\bar{\phi}_{n}^{2}-\beta\bar{\phi}_{n}\bar{\phi}_{n+1}\right).
\end{equation}
Here we have defined 
\begin{eqnarray} \label{beta in terms of lattice spacing}
\beta &=& \frac{2}{2+a^{2}m_{f}^{2}}, \\
\label{enot in terms of lattice spacing} E_{0} &=& \frac{(2+a^{2}m_{f}^{2})^{1/2}}{a}.
\end{eqnarray}
It is interesting to note that (i) under the scaling transformation
\eqref{scaling transformation}, $E_{0}\rightarrow\xi E_{0}$ while
$\beta$ is scale invariant. Thus, in writing the Hamiltonian
\eqref{discretized hamiltonian in terms of rescaled variables}, we
have separated the scale invariant part of the Hamiltonian from the
scale dependent part (which appears solely in the factor $E_{0}$). 
(ii) The canonical transformations (\ref{rescaling the canonical variables})
are well-defined for all values of $a$ (including the continuum limit whereby $\beta \to 1$ and $E_0 \to \infty$).
(The above canonical transformations have been discussed by Botero and Reznik \cite{5} in a 
different context.)

Since the determinant of the covariance matrix and, hence, the entanglement entropy are invariant 
under the canonical transformations, we can express the determinant of the covariance matrix 
(\ref{determinant reduced state})  in terms of $\beta$ by pulling out a factor of $4/a^{2}$ from both the 
numerator and the denominator leading to 
\begin{eqnarray} 
\label{covariance matrix for discretized scalar field in terms of beta}
\text{Det}(\sigma_{red}) &=&
\frac{1}{4N^{2}}\sum_{i=0}^{N-1}\frac{1}{\sqrt{1-\beta+2\beta\sin^{2}\left(\frac{\pi
      i}{N}\right)}} \nonumber \\ &&
\sum_{j=0}^{N-1}\sqrt{1-\beta+2\beta\sin^{2}\left(\frac{\pi
    j}{N}\right)}.
\end{eqnarray}

This points to the fact that, for the purpose of evaluation of the entanglement entropy, instead of working 
with the full Hamiltonian (\ref{discretized hamiltonian in terms of rescaled variables}), it is sufficient to work with the following 
`effective' Hamiltonian 
\begin{equation} \label{hamiltonian without e0}
\bar{H}^{\rm (1D)}=\frac{1}{2}\sum_{n=0}^{N-1}\left(\bar{\pi}_{n}^{2}+\bar{\phi}_{n}^{2}-\beta\bar{\phi}_{n}\bar{\phi}_{n+1}\right),
\end{equation}
from which we identify the following normal mode frequencies
\begin{equation} \label{artificial frequency}
\bar{\omega}_{i}=\sqrt{1-\beta+2\beta\sin^{2}\left(\frac{\pi i}{N}\right)}.
\end{equation}
Note that $\bar{\omega}$ is related to the normal mode frequency $\omega$ defined in \eqref{normal mode frequencies for discretized scalar field} by
\begin{equation} \label{relation between the frequencies}
\omega_{i}=\frac{2}{a}\bar{\omega}_{i}.
\end{equation} 

Thus, the resultant entanglement entropy depends {\it only} on the scale-invariant parameter $\beta$. As pointed out earlier,  in the continuum limit $\beta$ is finite 
while $E_0$ diverges.  

Eq.~(\ref{covariance matrix for discretized scalar field in terms of beta}) gives 
insight into the real cause of the divergence. To understand this, 
let us first consider the case where $\beta \to 1$. For $i = 0$ 
mode, the effective normal mode frequency is $\bar{\omega}_0 = \sqrt{1 - \beta}$. 
This implies that in Eq.~\eqref{covariance matrix for discretized scalar field in terms of
  beta} the discrete sum over $i$ would diverge because of the appearance of (at
least) one term with zero in the denominator (the zero mode term). This
shows that the divergence is due to the zero mode of $\bar{\omega}$.
In fact, in this limit, large number of near-zero modes  {\it accumulate} leading to the vanishing of the 
effective normal mode frequencies and hence divergence of the entanglement entropy.

Another (heuristic) way to understand how the innocuous canonical transformation 
(\ref{rescaling the canonical variables}) identifies the divergence of the 
entanglement entropy due to the near-zero modes is to look at the 
effective Hamiltonian (\ref{hamiltonian without e0}). In the continuum limit, 
the last term on the right becomes 
$$\beta\bar{\phi}_{n}\bar{\phi}_{n+1} \quad \rightarrow \quad \bar{\phi}_{n}^{2}$$ 
and this term cancels the second term in the Hamiltonian (\ref{hamiltonian without e0}). 
Thus, in the continuum limit, there are large number of modes which effectively 
behave as free particles. In the context of quantum field theory, zero modes 
are excluded on the grounds that they are not normalizable and that they do not 
have particle interpretation.  Although the zero modes 
have no physical effects, they carry \cite{1989-Ford.Pathinayake-PRD} an undetermined, 
non-zero energy $\langle {\bar \pi}^2 \rangle/2$. This undetermined non-zero energy indicates the 
divergence of the entanglement entropy in $(1+1)$--dimensions. 

\section{Isolating the divergent contribution to entanglement entropy $\Delta S=S_{0}-S_{m}$}

Having identified the divergence of the entanglement entropy in the
continuum due to the presence of zero frequency modes, we still need
to isolate the divergent term from the finite terms. We start by taking
$N \rightarrow \infty$ limit in Eq. \eqref{covariance matrix for discretized scalar field
  in terms of beta}, and replace the summations by
integrals (with $\pi i/N\rightarrow\lambda$)
\begin{eqnarray} \label{covariance matrix for continuum in terms of beta}
{Det(\sigma_{red})}
&=&\frac{1}{\pi^{2}}\int_0^{\pi/2}\frac{d\lambda}{\sqrt{(1-\beta)+2\beta\sin^2(\lambda)}} \nonumber \\
&& \times\int_0^{\pi/2}\sqrt{(1-\beta)+2\beta\sin^2(\lambda)}d\lambda.
\end{eqnarray}
In the continuum limit $\beta\rightarrow 1$, and as noted earlier, the first integral
diverges. To isolate this divergence we introduce a cut-off function
$f_{\delta}(\lambda)=e^{-\delta(\pi/2-\lambda)/\lambda}$ which has the
desired property
\[
 f_{\delta}(\lambda)\rightarrow
     \begin{cases}
       \ 0 &\text{when $\lambda\rightarrow 0$}\\
       \ 1 &\text{$\lambda\rightarrow \pi/2$}.\
       \end{cases}
\]

From the first integral in \eqref{covariance matrix for continuum in
  terms of beta} it is clear that in the continuum ($\beta=1$) the
divergent contribution comes only from the region $\lambda\approx0$
where $\sin\lambda\approx\lambda$ and we get
\begin{equation} \label{isolating divergence}
\frac{1}{\pi}\left(\int_0^{\pi/2}\frac{1}{\lambda} e^{-\delta\frac{(\pi/2-\lambda)}{\lambda}}d\lambda\right)
=e^\delta\int_{\delta}^{\infty}t^{-1}e^{-t}dt=\frac{1}{\pi}e^{\delta}\Gamma(0,\delta).
\end{equation}
RHS of the above expression, in the limit when the cut-off parameter
$\delta\rightarrow0$, can be approximated by $(-\gamma-\log(\delta))$
where $\gamma\approx0.57$ is the Euler-Mascheroni constant.

The remaining terms in the series expansion of $\sin\lambda$ in the
first integral in \eqref{covariance matrix for continuum in terms of
  beta} give a finite contribution of $\log(4/\pi)/\pi$ while the
contribution from the second integral is $1/\pi$ (in evaluating these
finite contributions we have put $\delta=0$).

Finally, using \eqref{entropy of the reduced state} for $m=1$, we find
the entropy to be the sum of a finite contribution and a divergent
contribution
\begin{eqnarray} \label{entanglement entropy in 1+1 continuum}
S_{1} 
&=&\frac{1}{2}\log\left(\left(\frac{1}{\pi}\right)^2\left(\log(\frac{4}{\pi})-\gamma-\log(\delta)\right)\right)\\
&\approx&-\log(\pi)+\frac{1}{2}\log(\log(\frac{1}{\delta})).
\end{eqnarray}
It is interesting to note that the divergence in entropy is very slow
going only as a double logarithm. Using the Boltzmann definition of entropy 
as logarithm of number of states, Eq. (\ref{entanglement entropy in 1+1 continuum}) 
gives 
\begin{equation}
\Omega \sim  \left(\frac{1}{\pi}\right)^4 \left(\log\left(\frac{4}{\pi}\right)-\gamma - \log(\delta) \right)^2 
\end{equation}
The above expression explicitly shows that large number of near-zero modes ($\delta \to 0$) 
leads to the divergence of the entanglement entropy. 

Until now, we have focused on the entanglement in $1:N-1$ partition as entanglement 
entropy can be evaluated analytically.  It is straight forward to extend 
to a general partition $m:N-m$ using Eq. \eqref{entropy of the
  reduced state}. However, for a general partition it is no longer
possible to obtain analytic results and the determinant of the
covariance matrix \eqref{covariance matrix for discretized scalar
  field in terms of beta}, for instance, has to be computed
numerically.

\begin{figure}[ht!]
\centering
    \includegraphics[width=80mm]{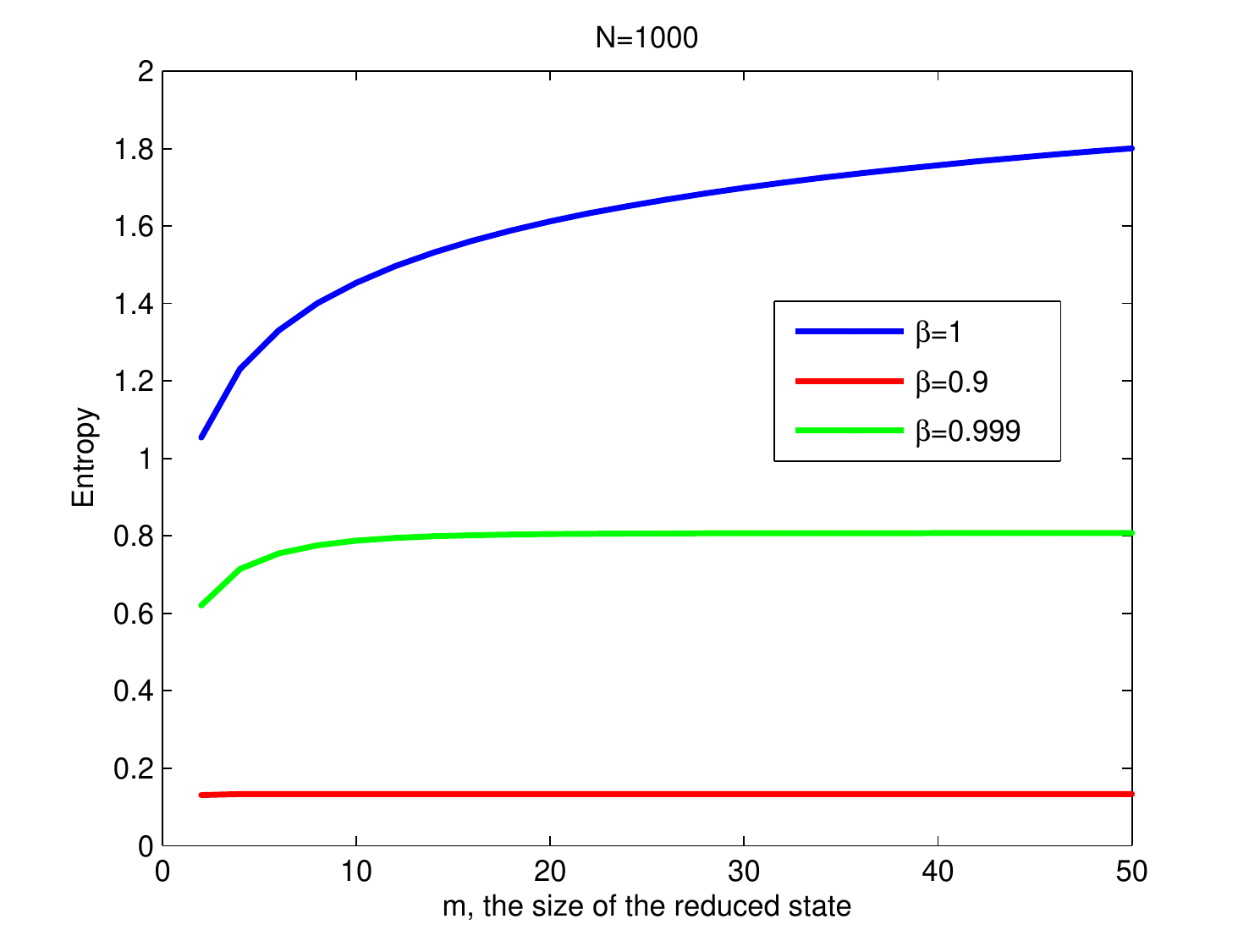}
  \caption{Entropy v/s the size of reduced state, with fixed boundaries. For $\beta = 1$, the area law is violated}\label{entropy in 1+1 dim as a function of m}
\end{figure}

Figure \ref{entropy in 1+1 dim as a function of m} shows the plot of
entropy as a function of $m$ (the number of oscillators traced over) 
for different values of $\beta$
and it is seen to be an increasing function of $m$.
Since entanglement entropy depends on the correlations across boundary
\cite{9}, one would expect that only oscillators near the boundary of
the traced out region would contribute which would imply that the
entropy should be a constant in $(1+1)$ dimensions even as more and more
oscillators away from the boundary get traced out. In one dimensional
chains, although this holds for weakly coupled chains as shown by
Plenio et.al ~\cite{6}, it breaks down in the strongly coupled
systems~\cite{srednicki} and the entropy grows as $\log(m)$.
\ \\

For the critical case of $\beta=1$, introducing the cut off function
$f_{\delta}$ defined earlier, the entropy $(S)$ for $m$ oscillator
reduced state gives a numerical fitting (ignoring the terms vanishing
in $\delta\rightarrow0$ limit)
\begin{eqnarray} \label{numerical fit entropy 1+1 dim}
S_{m}=C_0\log(m)+\frac{1}{2}\log\left(\log\left(\displaystyle\frac{1}{\delta}\right)\right)-\log(\pi),
\end{eqnarray}  
where $C_{0}$ is a constant.

\section{Entanglement entropy in higher dimensions}

To investigate the effects of dimensionality on the behavior of
entanglement entropy, we first consider the scalar field in $(2+1)$
dimensions (discretized on a square lattice). The discretized version
of the Hamiltonian for a scalar field in $(2+1)$ dimensions is
\begin{eqnarray} \label{discretized hamiltonian 2+1 dim}
H^{\rm (2D)} &=& \sum_{i,j =1}^{N} \frac{P(i,j)^2}{2a^2}+\frac{1}{2}a^2 m_f^2\eta(i,j)^2  \\
& &+\frac{1}{2}\left[\eta(i+1,j)-\eta(i,j)\right]^2  +\frac{1}{2}\left[\eta(i,j+1)-\eta(i,j)\right]^2 .\nonumber
\end{eqnarray}
Here $\eta(i,j)$ denotes the oscillator at position labeled by
$(i,j)$ on the two dimensional lattice and $P(i,j)$ is the
corresponding conjugate momentum. The Hamiltonian can be diagonalized
by going to normal coordinates and the corresponding normal mode
frequencies are (compare with Eq. \eqref{normal mode frequencies for
discretized scalar field} for $(1+1)$ dimensions)
\begin{eqnarray} \label{normal mode frequency 2+1 dim}
\omega(i,j)=\sqrt{m_f^2+\frac{4}{a^2}\left(\sin^2\left(\displaystyle\frac{\pi i}{N}\right)+\sin^2\left(\displaystyle\frac{\pi j}{N}\right)\right)}
\end{eqnarray}
where $(i,j =0,1,\dots,N-1)$.

As in $(1 + 1)-$ dimensional case,  the covariance matrix for the
single oscillator reduced state is given by
\begin{equation} \label{covariance matrix single oscillator 2+1 dim}
\sigma_{red} =\frac{1}{2N^2} \left[ \begin{array}{cc}
\sum\limits_{k} \sum\limits_{l} \frac{1}{a^{2}\omega(k,l)} & 0 \\
0 &\sum\limits_k\sum\limits_l a^{2}\omega(k,l) \end{array} \right] \, .
\end{equation}
The position and momentum covariance corresponding to the ground state of the 
Hamiltonian (\ref{discretized hamiltonian 2+1 dim}) are 
{\small
\begin{eqnarray} 
\label{position covariance 2+1 dim}
\Delta\eta &=& \frac{1}{a}\frac{1}{2N^{2}}\sum_{i,j}
\frac{1}{\sqrt{a^{2}m_f^2+4\left[ \sin^2\left(\displaystyle\frac{\pi i}{N}\right)+\sin^2\left(\displaystyle\frac{\pi j}{N}\right)\right]}} ~~~\\
\label{momentum covariance 2+1 dim}
\Delta P &=& \frac{a}{2N^{2}}\sum_{i,j}
\sqrt{a^{2}m_f^2+4\left[\sin^2\left(\displaystyle\frac{\pi i}{N}\right)+\sin^2\left(\displaystyle\frac{\pi j}{N}\right)\right]} ~~~~
\end{eqnarray} 
}
Comparing the continuum limit of the above expressions with that of Eqs. (\ref{position covariance 1+1 dim},\ref{momentum covariance 1+1 dim}) we notice that: 
(i) the position covariance diverges in both the cases,
(ii) the momentum covariance vanishes in $(2 + 1)-$dimensions while it is a constant in $(1 + 1)-$dimensions.
(iii) the determinant of the covariance matrix, remains finite in $(2 + 1)-$dimensions while it divergences in $(1 + 1)-$dimensions. 

To explicitly see the finiteness of the determinant, let us replace the summations in (\ref{position covariance 2+1 dim}, \ref{momentum covariance 2+1 dim}) with 
integrals (with $\theta\equiv\pi i/N$)
\begin{eqnarray} \label{det covariance matrix 2+1 dim}
&& Det(\sigma_{red})=\frac{4}{(\pi)^4}\int_0^{\pi/2}\int_0^{\pi/2}\displaystyle\frac{d\theta_1 d\theta_2}{\displaystyle\sqrt{\left(\sin^2\left(\theta_1\right)+\sin^2\left(\theta_2\right)\right)}} \nonumber \\
&& \times\int_0^{\pi/2}\int_0^{\pi/2}\sqrt{\sin^2\left(\theta_1\right) +\sin^2\left(\theta_2\right)}d\theta_1 d\theta_2.
\end{eqnarray}
Notice that, unlike $(1+1)$ dimensions, the result is finite as
can easily be seen by replacing $\sin\theta\approx\theta$ (regions
away from $\theta=0$ not contributing to the divergence). 

The entanglement entropy is given by \eqref{entropy of the reduced state}
%
%
and as in $(1 + 1)-$dimensional case, it is invariant under the scaling 
transformations: 
\begin{eqnarray} 
\label{scaling transformation2}
m_{f} &\rightarrow& \xi m_{f} \nonumber \\
a &\rightarrow& \xi^{-1}a,
\end{eqnarray}
and in the continuum limit,  the entropy is divergent due to the UV modes.

Taking into account the scaling symmetry, introduce a canonical rescaling (similar to that in $(1+1)-$dimensions \eqref{rescaling the canonical variables})
\begin{eqnarray} \label{scaling transformation 2+1 dim}
\bar{P}(i,j) &=& \displaystyle\frac{P(i,j)}{\left(a^2(4+a^2m_f^2)\right)^{\frac{1}{4}}},\nonumber \\
\bar{\eta}(i,j) &=& \eta(i,j)\left(a^2(4+a^2m_f^2)\right)^{\frac{1}{4}}.
\end{eqnarray}
In terms of the transformed variables the Hamiltonian \eqref{discretized hamiltonian 2+1 dim} becomes
\begin{eqnarray} \label{transformed hamiltonian 2+1 dim}
H^{\rm (2D)} &=& \frac{E_{0}}{2}\sum_{j=1}^N\sum_{i=1}^{N}(\tilde{P}(i,j)^2+\tilde{\eta}(i,j)^2-\beta\tilde{\eta}(i,j)\tilde{\eta}(i+1,j) \nonumber \\
&& -\beta\tilde{\eta}(i,j)\tilde{\eta}(i,j+1)),
\end{eqnarray}
where 
\begin{equation}
 E_{0}= \frac{\sqrt{4+a^2 \, m_f^2}}{a}; \qquad 
 \beta = \frac{2}{(4+a^2 \, m_f^2)}.
\end{equation}
It is important to highlight the differences between $(1 + 1)-$dimensions and higher dimensions: 
First, unlike Eqs. (\ref{scaling transformation}),  canonical transformations
(\ref{scaling transformation 2+1 dim}) in $(2 + 1)-$ dimensions have an explicit
$a$ dependence. This is due to the fact that the field and the canonically conjugate
momentum  $(\phi({\bf x}, t), \pi({\bf x}, t))$ have the dimensions 
$([L]^{-1/2}, [L]^{1/2})$. In the case of $(1 + 1)-$ dimensions $\phi({x}, t)$
($\pi({x}, t)$) are dimensionless. Hence, the strict $a = 0$ is ill-defined 
in the case of $(2 + 1)-$ (or higher) dimensions,

Second, like in $(1 + 1)-$dimensions, the Hamiltonian (\ref{transformed hamiltonian 2+1 dim}) is separated into a scale invariant part and a scale dependent 
part. In the small $a$ limit, $\beta \to 1/2$. 
Third, the position and the momentum covariance corresponding to the rescaled Hamiltonian (\ref{transformed hamiltonian 2+1 dim}) are
{\small
\begin{eqnarray}
& & \!\!\!\!\!\!\!\!\!\!\!\!\!\!   \Delta \eta =\frac{1}{2N^2}\sum_{l,m}\frac{1}{\sqrt{1-2\beta+2\beta\sin^2\left(\frac{\pi l}{N}\right)+2\beta\sin^2\left(\frac{\pi m}{N}\right)}}\\
& & \!\!\!\!\!\!\!\!\!\!\!\!\!\!  \Delta P =\frac{1}{2N^2}\sum_{l,m}\sqrt{1-2\beta+2\beta\sin^2\left[\frac{\pi l}{N}\right] 
+ 2 \beta \sin^2 \left[\frac{\pi m}{N}\right]}
\end{eqnarray}
}
are finite. This should be contrasted with the $(1+1)$ dimensional case 
where the canonical rescaling does not change the behavior of the position and momentum covariance.

Finally, like in $(1+1)$ dimensions,  the entropy depends \emph{only} on
$\beta$ and, hence, it is sufficient to work with the following effective 
Hamiltonian 
\begin{eqnarray} \label{hamiltonian 2+1 dim without e0}
\bar{H}^{\rm (2D)} &=& \sum_{j=1}^N\sum_{i=1}^{N}(\bar{P}(i,j)^2+\bar{\eta}(i,j)^2-\beta\tilde{\eta}(i,j)\tilde{\eta}(i+1,j) \nonumber \\
&&-\beta\tilde{\eta}(i,j)\tilde{\eta}(i,j+1)),
\end{eqnarray}
whose normal mode frequencies are
\begin{equation} \label{artificial frequency 2+1 dim}
\bar{\omega}(i,j)=\sqrt{1-2\beta+2\beta\bigg[\sin^{2}\left(\frac{\pi i}{N}\right)+\sin^{2}\left(\frac{\pi j}{N}\right)\bigg]}.
\end{equation}
Although the canonical transformations are not well-defined in $a \to 0$ limit, 
$\beta$ and normal mode frequencies are well-defined in this limit. 
The above calculation can be extended  to any higher space dimensions.  
The only difference is that the critical value of $\beta = 1/D$ where $D$ is the 
number of space dimensions. 

Heuristically, as in the previous case, in $a \to 0$ limit, the last three terms in the RHS of 
the Hamiltonian  \eqref{hamiltonian 2+1 dim without e0} cancel giving rise to the free 
particle Hamiltonian. This shows that there is an accumulation of near-zero modes, 
however, unlike the previous case the contribution is non-divergent. 

It is important to note that in both the cases [$(1 + 1)-$ and $(2 +1)-$dimensional 
space-time], in the continuum limit, the entanglement entropy behaves differently.  
One possible explanation for this difference is the following: Entanglement entropy arises due to the correlations across the
boundary \cite{9}.  In $(1+1)$ dimensions the boundary separating the
two regions is the zero dimensional point and a single oscillator,
occupying a point in space, is dense in the boundary (in the set-theoretic sense) leading to a divergent entropy.  On the other hand,
the boundary in $(2+1)$ dimensions is one dimensional and the single
traced over oscillator, occupying a point is not dense in the
1-dimensional spatial boundary and, hence, the entropy is finite.

\section{Conclusions}

Entanglement entropy for free fields is divergent. In this work
we revisited the question of the divergence in the light of zero 
modes. To obtain better analytic control we focused on the case where
only a single oscillator (in the discretized version of free scalar
field) was traced over. We found that the entanglement entropy 
has a scaling symmetry which the Hamiltonian does not possess. 
In terms of the canonically transformed variables, the Hamiltonian 
can be separated into a scale dependent and scale invariant part. 
We have shown that the origin of the divergence of the entanglement
entropy in $(1+1)-$dimensions in the continuum limit is due the presence of accumulation 
of large number of  (near-)zero frequency modes.  In the context of 
quantum field theory, zero modes are excluded on the grounds that 
they are not normalizable and that they do not have particle interpretation. 
Although the zero modes have no physical effects, they carry \cite{1989-Ford.Pathinayake-PRD}  an 
undetermined, non-zero energy $\langle {\bar \pi}^2 \rangle/2$. Using an IR cutoff, we have shown 
that the undetermined non-zero energy leads to the divergence of the 
entanglement entropy in $(1+1)-$dimensions.   

In higher dimensions, although, there is an accumulation of near-zero modes, 
their contribution to the entanglement entropy is non-divergent. One possible 
explanation for this difference between  $(1+1)$ and higher dimensions is the following: 
Entanglement entropy arises due to the correlations across the
boundary \cite{9}.  In $(1+1)$ dimensions the boundary separating the
two regions is the zero dimensional point and a single oscillator,
occupying a point in space, is dense in the boundary (in the set-theoretic sense) leading to a divergent entropy.  On the other hand,
the boundary in $(2+1)$ dimensions is one dimensional and the single
traced over oscillator, occupying a point is not dense in the
1-dimensional spatial boundary and, hence, the entropy is finite.

\section*{Acknowledgements}
We would like to thank A.P. Balachandran, Samuel Braunstein, Sourendu
Gupta, Namit Mahajan and V.P. Nair for useful discussions. KM is
supported by DST, Government of India through KVPY fellowship.  The work of SS and RT is supported by Max Planck partner group in India. SS is partly supported by Ramanujan Fellowship of DST, India. Work of TP is partly supported by J. C. Bose Fellowship of DST, India.

\end{document}